\documentclass[usegraphicx,useAMS,usenatbib]{mn2e}

\newcommand{\ulum}{{\rm erg}\, {\rm s}^{-1}}
\newcommand{\ucol}{{\rm cm}^{-2}}

\title[X-rays from a narrow-line RQ]{On the origin of the X-ray emission from a narrow-line radioquasar at $z>1$}
\author[X. Barcons et al.]
       {X. Barcons$^1$\thanks{E-mail: barcons@ifca.unican.es}, R. Carballo$^2$, F.J. Carrera$^1$, M.T. Ceballos$^1$
\newauthor J.I. Gonz\'alez-Serrano$^1$, J.M. Paredes$^3$\thanks{CER on Astrophysics, Particle Physics and Cosmology. Universitat de Barcelona}, M. Rib\'o$^3$, R.S. Warwick$^4$ \\
$^1$ Instituto de F\'\i sica de Cantabria (CSIC-UC), 39005 Santander, Spain\\
$^2$ Departamento de Matem\'atica Aplicada y Ciencias de la Computaci\'on, Universidad de Cantabria, 39005 Santander, Spain\\
$^3$ Departament d'Astronomia i Meteorologia, Universitat de Barcelona, Av. Diagonal 647, 08028 Barcelona, Spain\\
$^4$ Department of Physics \& Astronomy, University of Leicester, Leicester LE1 7RH}

\date{7 February  2003}

\pagerange{\pageref{firstpage}--\pageref{lastpage}}
\pubyear{2003}

\begin{document}

\maketitle

\label{firstpage}

\begin{abstract}
We present new $XMM-Newton$ X-ray observations of the $z=1.246$
narrow-line radioquasar RX~J1011.2+5545 serendipitously discovered by
$ROSAT$. The flat X-ray spectrum previously measured by $ROSAT$ and
$ASCA$ is shown to be the result of a steep $\Gamma\sim 1.8$ power law
spectrum seen through a moderate intrinsic absorbing column ($N_{\rm H}\sim
4\times 10^{21}\, \ucol$). The position of the X-ray source is
entirely coincident with the nucleus of the radio source that we have
resolved in new sensitive $VLA$ observations at 3.6 and 6 cm, implying
that scattering in the radio lobes is not responsible for the bulk of
X-ray emission.  In the EPIC pn image, a faint patch of X-ray emission
is apparent 14'' to the NE of the main X-ray source. The former is
positionally coincident with an apparently extended optical object
with $R\sim 21.9$, but there is no associated radio emission, thus
ruling out the possibility that this represents a hotspot in a jet
emanating from the primary X-ray source.  No reflection features are
detected in the X-ray spectrum of the narrow-line radioquasar,
although an Fe line with equivalent width of up to 600 eV cannot be
ruled out.
\end{abstract}

\begin{keywords}
X-rays: galaxies, galaxies: active
\end{keywords}

\section{Introduction}

One of the most debated aspects of the unification scheme for Active
Galactic Nuclei (AGNs) is the origin of the difference between
Radio-Loud (RL) and Radio-Quiet (RQ) objects. Whether or not an AGN is
able to develop a large-scale jet must be somehow intimately related
to the central engine, for example, to the spin of the black hole
(Blandford \& Znajek 1977). X-ray observations are
particularly relevant to this issue, as X-rays are thought to arise
mostly in the neighbourhood of the nucleus itself.  

It was noted early on (Wilkes \& Elvis 1987, Canizares \& White 1989)
that the X-ray spectra of RL AGN are generally harder than those of RQ
AGN. Further, it was found that the circumnuclear environment of RL and RQ
AGNs appears different: RL AGNs are often surrounded by cold absorbing
material (see, e.g., Cappi et al. 1997, Sambruna et al. 1999) while many
RQ AGN exhibit X-ray warm absorber features (Reynolds \& Fabian
1995). 

The origin of the apparently flat X-ray spectrum of RL objects was
attributed to beamed emission, which would dominate over, if at all
present, the Compton upscattering by relativistic electrons off the UV
photons from the accretion disk in RQ AGN.  The apparently flatter
slope of the underlying unabsorbed spectrum in RL AGN (Cappi et al.
1997) lent some support to this. However, more recent work by
Sambruna, Eracleous \& Mushotzky (1999) demonstrates that the
intrinsic power law (once absorption is accounted for) is very similar
in RL objects with a double-lobe radio structure (i.e., not suspected
of being beamed towards the observer) and RQ AGN (photon spectral
index $\Gamma=1.8-2.0$).

Nevertheless there are differences in the X-ray emission properties of
RL and RQ AGN, especially in the features that are thought to arise in
the reprocessing of the primary radiation.  Both the Fe K$\alpha$
fluorescence line and the Compton reflection shoulder, ubiquitous in
RQ objects (Pounds et al. 1990, Nandra \& Pounds 1994), are typically
weaker in RL objects (Sambruna et al. 1999).  This probably implies a
different structure in the inner accretion disk in RL and RQ
objects. Mechanisms that suppress the reflected component include a
truncated accretion disk converted into a geometrically thick
structure (e.g. an ADAF, see Meier 2001), and reflection in an ionized
disk (Ballantyne, Ross \& Fabian 2002). Hasenkopf, Sambruna \&
Eracleous (2002) analyzed the $ASCA$ and $BeppoSAX$ X-ray spectra of 3
moderate-redshift radioquasars with luminosities $\sim 10^{45}\,
\ulum$ and found similar properties, i.e., presence of cold absorption and
weak reflection components.

It is intriguing that absorption by cold material appears to play a
role both in the distinction between RL and RQ objects as well as in
the optical emission line properties of AGN.  RQ narrow-line AGNs
(Seyfert 2s) tend to be more absorbed than RQ broad-line AGNs (Seyfert
1s and QSOs) as predicted by the unified AGN scheme (Antonucci
1993). However a number unabsorbed Seyfert 2 galaxies has been found (Pappa
et al. 2001, Panessa \& Bassani 2002), lending some support to the idea
that optical emission line properties are, at least in some cases,
intrinsic to the Broad Line Region rather than modulated by
reddening/absorbing material (Barcons, Carrera \& Ceballos 2003).

In this paper we study a narrow-line radioquasar (RX~J1011.2+5545) at
$z=1.246$, one of the few high redshift examples of this class. It was
discovered in the $ROSAT$ Medium Sensitivity Survey (Carballo et al.
1995), where its PSPC X-ray spectral shape qualified it as the hardest
of the sources in the survey. A subsequent follow-up (Barcons et al.
1998) revealed that its $R\sim 21.0$ optical counterpart is a RL
narrow-line QSO, where the CIII]$\lambda 1909$ semi-forbidden line has
no detectable broad components, and MgII $\lambda 2800$ is extremely
weak.  The combined $ROSAT$ and $ASCA$ X-ray spectrum suggested
intrinsic absorption (at a low significance level) and also hinted an
underlying flat X-ray spectrum. Barcons et al. (1998) derived an X-ray
luminosity of $2.4\times 10^{44}$ and $1.1\times 10^{45}\, {\rm erg}\,
{\rm s}^{-1}$ in the $0.5-2$~keV and $2-10$~keV bands respectively
($H_0=70\, {\rm km}\, {\rm s}^{-1}\, {\rm Mpc}^{-1}$, $\Omega_m=0.3$
and $\Omega_\Lambda=0.7$ used throughout). The double-lobed morphology
of the radio counterpart ensures that beamed emission does not
dominate the X-ray spectrum. This source is therefore an interesting
example of an AGN that {\it should} be absorbed by cold gas, both
because it is radio loud and because its optical spectrum exhibits
very weak (if at all) optical broad emission lines.

\section{$XMM-Newton$ X-ray observations}

RX~J1011.2+5545 was observed for about 33 ks by $XMM-Newton$ (Jansen et
al 2001) on the 23 of November of 2001 during revolution 353, within
the AO-1 programme. All the instruments were on, although the prime
objective, given the flux of the source, was to use the EPIC cameras
(Turner et al. 2001, Str\"uder et al. 2001). All the EPIC cameras were
operated in full window mode and equipped with the 'Thin1' filter. The
observation data file was pipeline-processed by SAS v5.2, but the
analysis presented here has been conducted entirely using SAS v5.3.3.

The EPIC event lists (MOS1, MOS2 and pn) were cleaned of high
background flares, resulting in good-time intervals of 29.7, 30.0
and 24.6 ks respectively.  Further filtering followed the standard
procedures, keeping only single and double events and, in the case of
EPIC pn, those with {\tt FLAG=0}.  

Since the X-ray data has to be compared to much better angular
resolution optical and radio data, we also refined the astrometric
solution of the pipeline event lists and images which is based on the
attitude and orbit control system of the $XMM-Newton$ spacecraft.  This
astrometric solution is known to leave small (up to a few arcsec) but
significant residuals.  We then extracted all sources in the USNO-A2
catalogue in the EPIC field of view, and looked for a shift and
rotation in the EPIC pn source list provided by the pipeline by using
the internal SAS task {\tt eposcorr}.  A total of 47 X-ray to USNO-A2
matches were found, and the correction that we applied to the X-ray data was
$\Delta\alpha= 0.88''E$, $\Delta\delta=1.02''S$ and a rotation of
$-0.4^{\circ}$. 

\subsection {X-ray morphology}

\begin{figure}
\includegraphics[width=8cm]{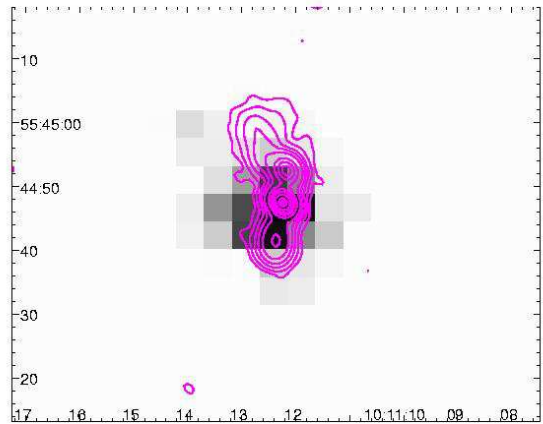}
\caption{$XMM-Newton$ EPIC pn image of RX~J1011.2+5545 in the
$0.2-12$~keV band, shown in equatorial coordinates (J2000). The pixel
size is $4''\times 4''$. The grey scale is linear with values between
10 and 100 counts per pixel. Contours are from the 3.6 cm radio data
with spacing defined under caption of Fig.~\ref{VLAimage}}
\label{Ximage}
\end{figure}

Our narrow-line radioquasar is well detected in all 3 EPIC
cameras. Furthermore, in the EPIC pn and in the EPIC MOS1 cameras, the
source shows some structure, in terms of X-ray emission towards the
NE. Figure~\ref{Ximage} shows the EPIC pn image in the $0.2-12$~keV band
around the target, where the NE emission is most obvious. The MOS1
image shows a similar structure, but this NE emission is not present
in the MOS2 data.  We attribute this fact to the ``triangular'' shape
of the point spread function (PSF) of the X-ray telescope corresponding to
the MOS2 detector, which is highly asymmetric.

The pipeline processed source list for these data does not find a
second source in the position of the NE extension, which lies
approximately 14'' away from the central source.  We tried hard to
modify the source finding algorithm parameters, lowering the detection
likelihood to 6 (instead of 10) and enabling 2 sources to be fitted
around each detection box.  However, the SAS refused to find a source
there.  A test for extended emission only gave a marginal detection
(likelihood $\approx 6$) in the MOS1 image.  

In spite of this, it is clear from Fig.~\ref{Ximage} that the X-ray
emission is real.  No other X-ray sources in the image show a similar
patch in the same direction, thus ruling out a bad attitude solution
origin for this second source. Our extracted spectrum for this source
contains $\sim 100$ and $\sim 50$ background-subtracted counts in the
EPIC pn and EPIC MOS1 images respectively (see sect.~2.2). 
Further support for the reality of this source comes from the
positional coincidence of this X-ray emission with an extended optical
source (see section 3). In what follows we will refer to this faint
X-ray source as ``source 2'', as opposed to ``source 1'', which is the
target object RX~J1011.2+5545.

\subsection {X-ray spectra}

X-ray spectra have been extracted from the 3 EPIC cameras, first for
the full emission complex (i.e., sources 1 and 2 together), as was
done for the $ROSAT$ and $ASCA$ data.  Spectra were extracted around
the centre of the target within a circle of 25 arcsec, resulting in
$\sim 1400$ background-subtracted counts for the EPIC pn spectrum and
over $\sim 500$ background-subtracted counts in each of the EPIC MOS
spectra. X-ray spectra were grouped in 10 count bins and counts below
$0.2$~keV were ignored.  No attempt was made to co-add spectra from
different instruments, but instead they were fitted together.

In the Barcons et al. (1998) analysis of the $ROSAT$ and $ASCA$
spectrum of this source, it was suggested that the X-ray spectrum
could be intrinsically flat ($\Gamma=1.4$ was the best fit) with some
marginal evidence for absorption (99\% level). We therefore fitted the
$XMM-Newton$ spectrum with a single power law plus galactic absorption
(fixed at $N_{\rm H}=7.8\times 10^{19}\, \ucol$).  The fit gave a very
poor $\chi^2=300.16$ for 229 degrees of freedom, with the residuals
immediatly suggesting the need for a further absorption component.
Adding an intrinsic cold absorber at the redshift of the source gives
a spectacular improvement with $\chi^2=194.72$ for 228 degrees of
freedom.  The F-test gives a significance of $3\times 10^{-23}$ for
absorption not being detected.  The best-fit parameters are reported
in table~\ref{Xraypars}.

\begin{table*}
\label{Xraypars}
\begin{center}
\caption{Parameters of the X-ray spectral fitting to the XMM-Newton
  EPIC data (see text for details). All errors are at 90\% confidence
  level for one parameter.}
\begin{tabular}{c c c c}
\hline
          & Sources 1+2             & Source 1               & Source 2 \\
\hline
$\Gamma$  & $1.83^{+0.07}_{-0.10}$  & $1.82^{+0.12}_{-0.12}$ &
          $1.94^{+0.23}_{-0.22}$\\
$N_H^a$   & $(4.1^{+0.7}_{-0.9})\times 10^{21}$ &
          $(4.2^{+1.4}_{-1.1})\times 10^{21}$ & --\\
Flux(0.5-2 keV)$^b$ & $6.9\times 10^{-14}$ & $6.7\times 10^{-14}$ &
          $\sim 10^{-14}$ \\
Flux(2-10 keV)$^b$ & $1.27\times 10^{-13}$ & $1.24\times 10^{-13}$ &
          $\sim 10^{-14}$ \\
$\chi^2$/d.o.f. & 194.72/228 & 110.55/109 & 7.75/13 \\
\hline
\end{tabular}
\end{center}
$^a$ Intrinsic absorbing column at $z=1.246$ in cm$^{-2}$\\
$^b$ Corrected for Galactic absorption in units of ${\rm erg}\, {\rm cm}^{-2}\, {\rm s}^{-1}$
\end{table*}

%(90\%
%confidence errors for a single parameter) are
%$\Gamma=1.83_{-0.10}^{+0.07}$ for the photon spectral index and
%$N_{\rm H}=(4.1_{-0.9}^{+0.7})\, \times 10^{21}\, \ucol$ for the intrinsic
%absorbing column at the redshift of the source $z=1.246$.

\begin{figure}
\includegraphics[height=8cm,angle=270]{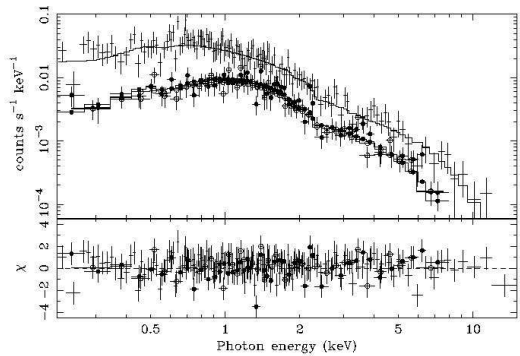}
\caption{$XMM-Newton$ EPIC (MOS1: filled circles, MOS2: hollow circles
and pn: error bars only) spectrum of the source 1 and 2 complex in
RX~J1011.2+5545, together with the best fit model (see text for
details). The bottom panel shows the contributions of individual bins
to the dispersion.}
\label{Xspectot}
\end{figure}

\begin{figure}
\includegraphics[height=8cm,angle=270]{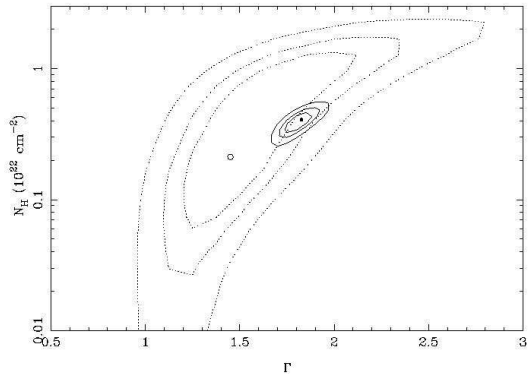}
\caption{Best fit (full point) and confidence contours in parameter
space (1,2, and 3 sigma) derived from the $XMM-Newton$ data (EPIC
MOS1+MOS2+pn) from the whole RX~J1011.2+5545 source (solid curves). For
comparison the hollow circle and the dotted curves show the best fit
and 1,2 and 3 sigma contours from the Barcons et al. (1998) analysis of the
$ROSAT$ and $ASCA$ data.}
\label{Xcont_tot}
\end{figure}

Figure~\ref{Xspectot} shows the measured X-ray (MOS1, MOS2, pn)
spectra and the corresponding best fit models, together with a plot of
the fitting residuals. Fig.~\ref{Xcont_tot} presents the confidence
contours in the ($\Gamma$,$N_{\rm H}$) parameter space, using the same
parameter range as employed by Barcons et al. (1998) in their study
based on $ROSAT$ (18 ks) and $ASCA$ (52 ks) spectra (solid curves for
XMM-Newton and dotted curves for $ROSAT$ and $ASCA$).  Besides
highlighting the excellent capability of $XMM-Newton$ to perform X-ray
spectroscopy of faint X-ray sources, Fig.~\ref{Xcont_tot} also shows
that all data sets are consistent.  The $XMM-Newton$ data, however,
clearly rule out an intrinsically flat X-ray spectrum, and show that
the spectral flatness inferred from the earlier data was due to a
combination of a steep power law seen through a significant absorbing
column.

We also attempted to fit an emission feature to the full 3 instrument
X-ray spectra, at the expected position of the redshifted Fe
K$\alpha$ line.  Hasenkopf et al. (2002) find line equivalent widths
around $100-200$~eV in 3 radioquasars of similar X-ray luminosity. However,
the residuals to the absorbed power-law fit to RX~J1011.2+5545 do not
show any hint of a significant positive deviation around $3$~keV, and
indeed no improvement in the $\chi^2$ is achieved by adding a gaussian
line.
%Figure~\ref{contFeline} shows the
%contours in the line intensity versus line width parameter space,
%resulting from a constrained parameter search which limited the line
%position between $6$ and $7$~keV (rest-frame) and the line dispersion
%between $0.1$ and $1$~keV. 
The 3$\sigma$ upper limit to the line
equivalent width is $\sim 1$~keV, and slightly lower ($\sim 600$~eV)
assuming a line width $<0.3$~keV. We have also tried to add a
reflection component (via the {\tt pexrav} model from Magdziarz \&
Zdziarski 1995), but the $\chi^2$ does not improve and returns a null
value for the reflected component.

%\begin{figure}
%\includegraphics[height=8cm,angle=270]{contfeline.ps}
%\caption{Confidence contours in parameter space (1,2, and 3 sigma)
%for the Fe line intensity (restricted to $6$-$7$~keV in the rest-frame)
%versus line dispersion.}
%\label{contFeline}
%\end{figure}

The fluxes (corrected for Galactic absorption) of the overall complex
are shown in table~\ref{Xraypars}.
%$6.9\times 10^{-14}$, and $12.4\times 10^{-14}\, {\rm erg}\, {\rm
%cm}^{-2}\, {\rm s}^{-1}$ in the $0.5-2$~keV and $2-10$~keV bands
%respectively.  
Note that these are entirely consistent with the $ROSAT$ and $ASCA$
inferred fluxes, i.e., the source has not varied much after an
interval of several years. The luminosity of the source is $2.5\times
10^{44}$ and $9.4\times 10^{44}\, \ulum$ in the 0.5-2 and 2-10 keV
bands respectively. If we correct for the intrinsic absorption, these
luminosities go up to $6.4\times 10^{44}$ and $9.8\times 10^{44}\,
\ulum$.

\begin{figure}
\includegraphics[height=8cm,angle=270]{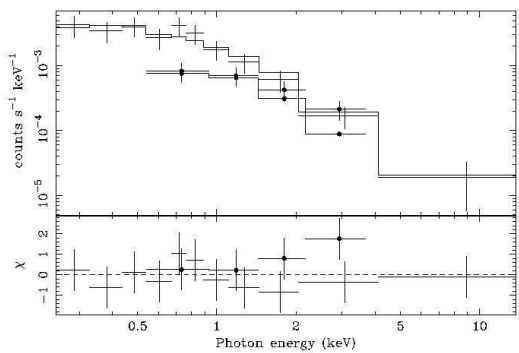}
\caption{$XMM-Newton$ EPIC (MOS1: filled circles and pn: error bars
only) spectrum of source 2 together with the best fit model (see text
for details). The bottom panel show the contributions of individual
bins to the dispersion.}
\label{Xspec_source2}
\end{figure}

In the EPIC MOS1 and pn data, sources 1 and 2 can be separated (see
again Fig.~\ref{Ximage}) and therefore we have extracted the spectrum
of sources 1 and 2 independently from these data.  In order to
minimize the contamination of the tail of source 1 on the spectrum of
source 2, conservative non-overlapping circles of radii 9.5~arcsec and
7.5~arcsec were choosen around the centre of sources 1 and 2
respectively. This gave $\sim 900$ pn and $\sim 315$ MOS 1
background-subtracted counts for source 1 and $\sim 100$ pn and $\sim
50$ MOS1 background-subtracted counts for source 2. The spectra of
source 1 is virtually equivalent to the overall spectrum, as it
contains more than 90\% of the total flux (see best fit parameters in
table~\ref{Xraypars}.  
%The best fit photon
%spectral index and intrinsic absorbing column are
%$\Gamma=1.82_{-0.12}^{+0.12}$ and $N_{\rm H}=(4.2_{-1.1}^{+1.4})\times
%10^{21}\, \, \ucol$, i.e., the same as in the fit to the overall
%complex, but with slightly larger errors.

The MOS1+pn spectrum of source 2 has been fitted to a power law model
with Galactic absorption, and the resulting parameters are shown also
in table~\ref{Xraypars}. Figure~\ref{Xspec_source2} shows the
X-ray spectrum together with the best fit model. 

\section{$NOT$ optical observations}

Given the interesting morphology of the EPIC X-ray image, the region around
RX~J1011.2+5545 was imaged in the optical with the 2.6m $NOT$ optical
telescope in La Palma.  The instrument ALFOSC was used with the
Johnson R filter, for a total integration of 1 hour (splitted in two
30 min integrations slightly shifted) in service time on the 24th of
October 2002. The observation was conducted in bright time and the
average airmass was 1.6. 

The images were reduced using standard IRAF
reduction techniques, including debiasing, trimming, flat-fielding
(using twilight flats) and finally registering both images to a common
frame.  The measured seeing in the resulting image is
1.37''. The astrometric calibration was done using the USNO-A2
catalogue. No photometric standard was observed, so we used the less
sensitive image of Barcons et al. (1998) to perform the photometric
calibration. 

\begin{figure}
\includegraphics[width=8cm]{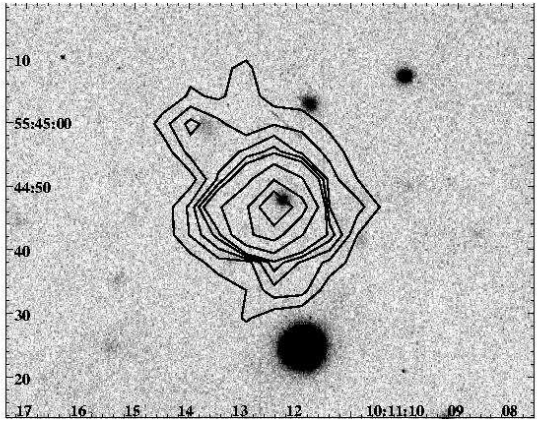}
\caption{$NOT$ R-band image of the RX~J1011.2+5545 region with
contours from the $0.2-12$~keV EPIC pn image overlayed. Contour levels
correspond to 8, 13.5, 19, 24.5, 30, 50, 80, 120 counts/pixel. }

\label{NOTimage}
\end{figure}

The $NOT$ image is shown in Fig.~\ref{NOTimage} along with contours
from the EPIC pn X-ray image.  It is then clear that source 1 is
coincident with the optical counterpart identified in Barcons et al.
(1998), with magnitude $R\sim 21.0$ and that source 2 overlays with a
fainter, extended optical source.  This optical counterpart has a FWHM
extent of 2.4~arcsec, while the seeing is 1.4~arcsec. We measure its
magnitude by aperture photometry to be
$R\sim 21.9$. This spatial coincidence lends further support to the
reality of source 2.

\section{$VLA$ radio observations}

The source RX~J1011.2+5545 is known to be a radio source, detected in
various surveys, as previously discussed in Barcons et al. (1998).  In
particular, the FIRST-VLA data at 1.4~GHz resolved the radio source into
two components, aligned approximately along the N-S axis, with a total
integrated flux of $\sim 0.16$~Jy. Furthermore, it was also detected
in the 7th Cambridge survey at 151 MHz as source 7C 434 with a flux of
1.2~Jy (Pooley, Waldram \& Riley 1998).

Besides the double-lobed morphology of the source, the FIRST-VLA radio
map shows a hint of extended emission from the northern lobe towards
the NE.  Since this is the general direction of source 2 detected
in the EPIC pn image, we requested a deeper, better angular resolution
observation of this source with the $VLA$ of the National Radio
Astronomy Observatory (NRAO) as an {\it ad hoc} proposal.

The observation was conducted in November 13, 2002 with the C
configuration.  We choosed both 3.6 cm (8.4~GHz) and 6 cm (5~GHz)
giving approximate beams of $\simeq 3$~arcsec and $\simeq 5$~arcsec
respectively for natural weigthing (see below). Exposure times were
selected in order to have a 3$\sigma$ sensitivity of approximately
0.1~mJy~beam$^{-1}$ at both frequencies. We recall that the FIRST-VLA survey
at 1.4~GHz has a flux threshold a factor of 5 higher and a beam of
$\simeq 5$~arcsec.  Since sources 1 and 2 are separated by 14'', the
radio maps (even the early FIRST-VLA survey) should be able to clearly
resolve both sources.  Moreover (see section 5) in the event that
source 2 corresponds to a knot in a jet emanating from source 1, a
flux level of the order of $\sim 1-10$~mJy would be expected, and in
this case a positive detection should be achieved with the obtained
radio data.

The observation consisted of two snapshots of 10 minutes at 3.6~cm and two
snapshots of 8 minutes at 6~cm on RX~J1011.2+5545, preceded and followed
by a 1.5 minute observation of the phase calibrator 1035+564. The
amplitude calibrators used were 3C~48 at 3.6~cm and 3C~286 at 6~cm (we were
forced to use different amplitude calibrators because of technical
problems during the 3.6~cm snapshot of 3C~286). The data were edited and
calibrated using standard procedures within the {\sc aips} software
package of NRAO. The maps shown here have been produced using
self-calibration and natural weighting of the data, which are the most
sensitive to extended and faint structures. The final 3$\sigma$ sensitivy
is 0.07~mJy~beam$^{-1}$ at 3.6~cm and 0.09~mJy~beam$^{-1}$ at 6~cm.

\begin{figure}
\includegraphics[width=8cm]{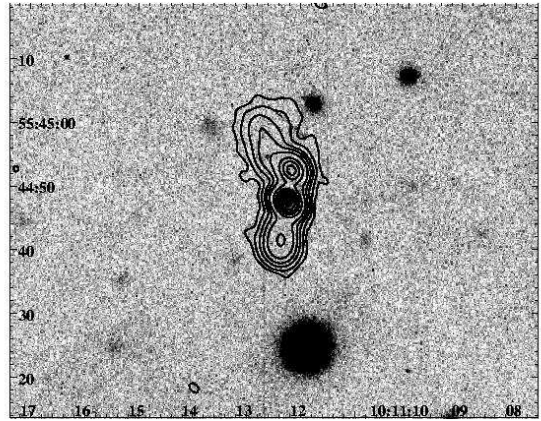}
\includegraphics[width=8cm]{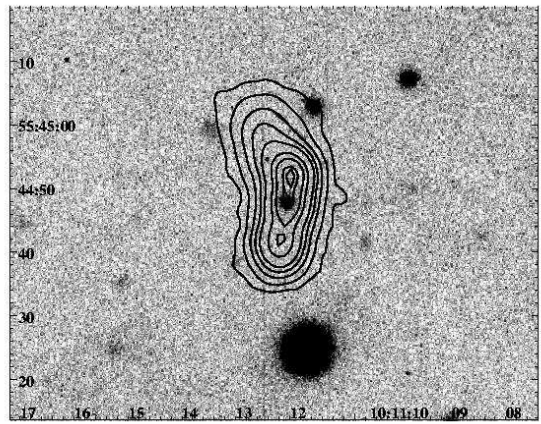}
\caption{Top: $NOT$ R-band image of the RX~J1011.2+5545 region with
contours from the $VLA$ 3.6 cm (X-band) data overlayed. Contour
levels correspond to 0.07, 0.2, 0.5, 1.0, 2.0, 5.0, 8.0, 11.0, 15.0,
20.0 and 22.0~mJy~beam$^{-1}$. Bottom: same image with contours from
the $VLA$ 6~cm (C-band) data overlayed, contour levels corresponding
to 0.1, 0.5, 1.5, 3.0, 5.0, 10.01, 14.3, 20.0 and
22.0~mJy~beam$^{-1}$. From top to bottom, we refer to the top NE tail
as ``Tail'', the northern lobe as ``N-lobe'', the central know
coincident with the optical counterpart as ``nucleus'' and the
southern lobe as ``S-lobe''}
\label{VLAimage}
\end{figure}

\begin{table}
\label{spectral-index}
\begin{center}
\caption{Spectral index and flux of the various components of the
radio source, computed from the clean components obtained when
cleaning of the raw radio maps}
\begin{tabular}{l c c c}
\hline
Component & $\alpha$ & 6 cm Flux & 3.6 cm Flux \\
          &          & (mJy)     &  (mJy)      \\
\hline
Tail & $-1: -1.5$    &  7.3      &   3.9 \\
N-lobe & $-0.8$      &  24.1     &   15.5\\
Nucleus & $+0.3$     &  19.9     &   23.7\\
S-lobe & $-1.2$      &  18.3     &    9.7\\
\hline
\end{tabular}
\end{center}
\end{table}

Fig.~\ref{VLAimage} shows the contours of the $VLA$ maps overlayed on the $NOT$
optical image. The 3.6 cm map clearly resolves the nucleus (concident
with the optical counterpart of RX~J1011.2+5545) and two lobes in the
N-S direction.  The 6 cm data exhibits the same structure, but without a
clear separation between the northern lobe and the nucleus. 

The variation in radio spectral index $\alpha$ (defined as
$S_{\nu}\propto\nu^{\alpha}$ between the 3.6 and 6 cm bands) across
the source ranges from $\alpha=-1.7$ to
$\alpha=+0.3$. Table~\ref{spectral-index} shows the radio spectral
index at various locations of the radio source (see caption of
Fig.~\ref{VLAimage} for definitions). The overall nucleus and
double-lobe structure is consistent with a synchrotron self-absorbed
nucleus plus ageing effects in the relativistic electron population
along the lobes.  The flatter spectral index of the N-lobe (as
compared to the S-lobe) could be due to a variety of reasons,
including the jet meeting a denser medium in the N-lobe zone,
different ordering of the magnetic field, the orientation of the
source etc. The steep-spectrum NE tail is consistent with freely
expanding radio gas.

%\begin{figure}
%\includegraphics[width=8cm]{spix.ps}
%\caption{Radio spectral index $\alpha$. The grey scale is linearly
%spaced between $\alpha=-2$ (white) to $\alpha=+0.3$ (black).}
%\label{spix}
%\end{figure}

It is also clear from Fig.~\ref{VLAimage} that source 2 detected in the
EPIC pn does not emit in radio, down to a 3$\sigma$ sensitivity below
$0.1$~mJy at both 3.6 and 6 cm.  To further emphasize this, we overlay
the 3.6 cm contours on the EPIC pn image in Fig.~\ref{Ximage} (see also
Fig.~\ref{NOTimage} for the X-ray contours on top of the optical
image). It can also be appreciated that, within the $XMM-Newton$
angular resolution, the X-ray emission matches very well that of the
nucleus resolved by the radio maps with little contribution from the
lobes.

\section{Discussion and conclusions}

\subsection{The nature of source 2}

Using the positions of the optical counterparts for sources 1
($\alpha_1=10:11:12.27$, $\delta_1=+55:44:47.9$, J2000) and 2
($\alpha_2=10:11:13.74$, $\delta_2=+55:44:59.9$, J2000) we can
compute the possibility of finding two such close sources by chance.
The relevant parameter is the source density at the flux of source 2
($0.5-2$~keV flux of $10^{-14}\, {\rm erg}\, {\rm cm}^{-2}\, {\rm
s}^{-1}$) which we take $\sim 140\, \deg^{-2}$.  The probability of
finding source 2 within 14 arcsec of source 1, being unrelated, is
$\sim 0.3\%$. Although this is not a very significantly small number,
we checked the possibility that source 2 could be somehow related to
source 1.

We have further explored the possibility that source 2 results from
the emission of a knot in a putative jet of source 1, should they be
at all related. (Note, however, that the geometry of the radio maps
does not suggest that hypothesis).  Such knots have been found to be
X-ray emitters in several objects, among them 3C273 (Marshall et al.
2001).  If we take the A1 knot as quoted in Marshall et al. (2001), and
scaling from the flux measured at $1$~keV from the $XMM-Newton$ data in
source 2 ($\sim 0.01-0.02\, \mu{\rm Jy}$) we would expect a radio flux
around 6 cm of $\sim 5-15$~mJy.  Our $VLA$ data, which sets a
3$\sigma$ upper limit of 0.1~mJy~beam$^{-1}$ at the position of source
2, clearly rules out that option.

We can therefore conclude that source 2 is a real X-ray source,
with an optical counterpart of $R\sim 21.9$ and to our knowledge
unrelated to source 1.

\subsection{The nature of the X-ray emission in the narrow-line
radioquasar}

The main conclusion of this work is that the X-ray emission properties
of this narrow-line radioquasar are very similar to those of
broad-line radioquasars of similar luminosity.  The apparently flat
X-ray spectrum found by $ROSAT$ and $ASCA$ is the result of a steep
power law (photon spectral index $\Gamma=1.83$) seen through an
intrinsic absorbing column $N_{\rm H}=4\times 10^{21}\, \ucol$, in agreement
with the results from Hasenkopf et al. (2002) for broad-line
radioquasars of similar luminosity.  The typically flatter spectral
index attributed earlier to RL AGNs $\Gamma \approx 1.4-1.5$ is
totally excluded in this narrow-line radioquasar (see again
Fig.~\ref{Xcont_tot}). This confirms that relativistic beaming does not
play a major role in its X-ray emission, as expected from the
double-lobe radio morphology.

The X-ray spectrum displays moderate cold absorption (in excess of
Galactic absorption), also in the same range as the Hasenkopf et al.
(2002) broad-line radioquasars of similar luminosity (but lower
redshift) $\sim (1-4)\times 10^{21}\, \ucol$. The overall spectral
energy distribution (shown in Barcons et al 1998) is similar to the
template for RL QSOs as shown in Fig. 10 of Elvis et al (1994), except
for the harder X-ray slope resulting from intrinsic absorption.

The measured absorption column is, however, on the lower side of that
expected for narrow-line AGN.  Indeed, in a large sample of
[OIII]-selected narrow-line RQ AGN (Seyfert 1.8, 1.9 and
mostly 2 galaxies), Risaliti, Maiolino \& Salvati (1999) found that
the majority of these sources are absorbed by columns $>10^{22}\,
\ucol$. This should be expected in the framework of the AGN unified
model, where the lack or weakness of optical broad emission lines
would be due to reddening by the same gas that absorbs soft
X-rays. Nevertheless, a number of examples of RQ narrow-line Seyferts
have been found with small or no cold absorbing column at all (Pappa
et al. 2000, Panessa \& Bassani 2002, Barcons et al. 2003).  The
narrow-line radioquasar studied here also belongs to this class of
low-absorption column objects, where the lack of optical broad-lines
is difficult to understand in terms of reddening/absorption and could
be attributed to intrinsic properties of the broad line region.

It is also remarkable that no reflection features are present in this
object, down to the sensitivity level of our rather high quality X-ray
data. However the 3$\sigma$ upper limit found for the equivalent
width of the Fe line ($600$ eV in the case of a relatively narrow line
$\sigma<0.3$~keV) is not in conflict with a reflection component
consistently small, as in broad-line radioquasars of similar
luminosity (Hasenkopf et al. 2002).

It would be interesting to study other narrow-line radioquasars of
similar luminosity to check whether the properties of RX~J1011.2+5545
are peculiar (in the sense of moderate absorbing column of cold gas
and weak reflection features) or relatively general.  Bright and medium
sensitivity X-ray surveys performed with $XMM-Newton$ should be
particularly sensitive to this type of sources (even more heavily
absorbed), where modest follow-up exposures could deliver X-ray data
of enough quality to merit a spectral analysis as the one presented here.

\section*{Acknowledgments}
The work reported herein is based partly on observations obtained with
$XMM-Newton$, an ESA science mission with instruments and
contributions directly funded by ESA member states and the USA
(NASA). The $NOT$ telescope is operated by the Nordic Optical
Telescope Scientific Association on the spanish Observatorio del Roque
de los Muchachos of the Instituto de Astrof\'\i sica de Canarias.  We
are grateful to the service support for conducting the optical
observations.  The National Radio Astronomy Observatory is a facility
of the National Science Foundation operated under cooperative
agreement by Associated Universities, Inc. We acknowledge financial
support by the Ministerio de Ciencia y Tecnolog\'\i a (Spain), under
grants AYA2000-1690 (XB, FJC, MTC), AYA2002-03326 (RC, JIGS) and
AYA2001-3092 (MR, JMP). MR and JMP acknowledge also partial support by the
European Regional Development Fund (ERDF/FEDER). During this work, MR
has been supported by a fellowship from CIRIT (Generalitat de
Catalunya, ref. 1999~FI~00199).

\bsp

\label{lastpage}

\end{document}